# Flexible FPGA ECDSA Design with a Field Multiplier Inherently Resistant against HCCA

Zoya Dyka, Dan Kreiser, Ievgen Kabin and Peter Langendoerfer
*IHP*
*Im Technologiepark 25*
*Frankfurt (Oder), Germany*

*Abstract*— In this paper we describe our flexible ECDSA design for elliptic curve over binary extended fields *GF($2^l$)*. We investigated its resistance against Horizontal Collision Correlation Attacks (HCCA). Due to the fact that our design is based on the Montgomery *kP* algorithm using Lopez-Dahab projective coordinates the scalar *k* cannot be successful revealed using HCCA, but this kind of attacks can be helpful to divide the measured traces into parts that correspond to processing of a single bit of the scalar *k*. The most important contribution of this paper is that our flexible field multiplier is resistant against horizontal attacks. This inherent resistance makes it a valuable building block for designing unified field multipliers.

*Keywords*— *ECDSA; Montgomery kP algorithm; FPGA implementation; horizontal side channel analysis (SCA) attacks; flexible field multiplier; power traces; electromagnetic traces.*

I. INTRODUCTION

Elliptic Curve Cryptography (ECC) provides the same level of security as RSA based approaches with significantly shorter keys. This is especially needed in applications in which resource constraint devices are used e.g. Internet of Things, WSN, automation industry, protection of critical infrastructures. ECC can guarantee confidentiality of communication and can also be used for authentication of persons/devices. One of the potential applications is for example to ensure data integrity and authenticity of communication between cars and road side units (RSUs). Currently traffic lights can be switched without any protection means in place, the use of digital signatures based on ECDSA would prevent such misuse as it allows to determine whether or not the sending device has the proper access rights.

For signature generation corresponding to Elliptic Curve Digital Signature Standard (ECDSA) [1] each RSU has to use its ECC private key. According to Kerckhoff's principle the private key has to be kept secret while the applied cryptographically strong cipher algorithm, its input and output values may be publicly known. The goal of attackers is to reveal the private key. Cryptographic algorithms are implemented either in SW or in HW, i.e. they run on a device. The current through the device, its electromagnetic radiation and other physical parameters are caused by each execution of cryptographic operations. If an attacker has physical access to the device, for example to an RSU, he can exploit those side channel effects to reveal its private key. Misusing the key can cause dangerous situations in the traffic, car crashes, etc. In order to avoid malicious attacks the cryptographic implementations need to be protected against a wide variety of side channel analysis (SCA) attacks such as vertical, horizontal, differential, correlation, collision-based attacks etc.

Collision-based attacks [2]- [4], classical differential power analysis [5] and correlation power analysis attacks [6] are vertical attacks. Examples of horizontal attacks are: simple power analysis attacks e.g. [7], simple electromagnetic analysis attacks e.g. [8], the Big Mac attack [9], the localized electromagnetic analysis attack [10], horizontal collision correlation analysis attacks [11] - [13], horizontal differential power analysis (DPA) and differential electromagnetic analysis (DEMA) attacks [14], [15].

Vertical DPA attacks need more measurements than horizontal attacks to be successful exploiting the fact that there is some kind of relation between the different but known processed inputs and the always constant key. Therefore, well-known countermeasures such as EC point blinding, randomization of projective coordinates of EC points or the key randomization proposed in [16] are effective against vertical attacks. But these countermeasures are not effective against horizontal attacks. Thus, protection against vertical DPA is useful only if the design is well protected against horizontal DPA.

Elliptic Curve point multiplication with a scalar, denoted usually as *kP* operation, is the main operation for ECC. *P* is a point of the selected EC and has two affine coordinates: *P*=(*x, y*). The scalar *k* is a big binary number; it is a private key if the decryption corresponding to the ElGamal approach [17] is performed. For signature generation the scalar *k* is a random number. If an attacker can reveal the random scalar *k* used for a signature generation he can calculate the private key of the signer. As a consequence, the identity of the signer can be stolen. This means that implementations of the *kP* operation have to be resistant against SCA attacks. Designing a flexible accelerator for EC point multiplication, i.e. one that can be used for at least two different ECs for example over extended binary fields $GF(2^l)$, is a not trivial task, especially if the design has to be resistant against horizontal SCA attacks.

The main contributions of this paper are:
- performing a HCCA against our flexible ECDSA design;
- evaluation of the resistance of our flexible field multiplier against HCCA showing its inherent resistance.

The rest of the paper is structured as follows. In section II we give a short overview of the ECDSA protocol. Our flexible ECDSA design is described in section III and the HCCA performed is given in section IV. The resistance of our flexible field multiplier is examined in section V. The paper finishes with short conclusion.

## II. BACKGROUND: ECDSA

Elliptic Curve Digital Signature Algorithm (ECDSA) is an algorithm to generate digital signatures as described in the NIST documentation [1]. A generation and verification of signature can be implemented as a sequence of mathematical operations in finite fields. The signature generation and verification algorithms are sketched in TABLE I.

TABLE I. GENERATION AND VERIFICATION OF SIGNATURE CORRESPONDING TO ECDSA

| Alice has the ECC key pair ($Pub_A$; $key_A$) | Bob knows Alice's public key $Pub_A$ |
|---|---|
| Signature generation | Signature verification |
| 1. writes a message<br>2. $e$ = hash(message)<br>3. generates random $k < \varepsilon$<br>4. $T = (x_T, y_T) = k \cdot G$<br>5. $r = x_T \bmod \varepsilon$<br>6. $s = (e + r \cdot key_A)/k \bmod \varepsilon$<br>7. sends to Bob: (message, r, s) | 1. receives: (message, r, s)<br>2. $e$ = hash(message)<br>3. $u_1 = e/s \bmod \varepsilon$<br>4. $u_2 = r/s \bmod \varepsilon$<br>5. $T = (x_T, y_T) = u_1 \cdot G + u_2 \cdot Pub_A$<br>6. $r = x_T \bmod \varepsilon$ ? |

Corresponding to the ECDSA algorithm, public keys are points of an EC and private keys are big binary numbers, smaller than the order $\varepsilon$ of the base point $G$. $\varepsilon$ and $G$ are parameters of the EC that was selected for cryptographic operations; they are given in [1]. To generate the signature Alice calculates for the written message its hash $e$ of her message and performs the multiplication of the base point $G$ with a generated random scalar $k$ (see step 4 in TABLE I. ), i.e. an EC point multiplication, usually denoted as $kP$ operation. This operation is a sequence of mathematical operations in the finite field and it is the most time, energy and computation expensive part of the signing protocol. The result of the performed $k \cdot G$ operation is also a point of the EC. Its x-coordinate is part of digital signature, denoted as r. Using the calculated r and $e$, the generated scalar $k$ and the private key $key_A$ Alice calculates the last part of the signature, denoted as s (see step 6 in TABLE I. ). Alice sends her message to Bob with its digital signature, i.e. with two numbers: r and s.

Bob receives the signed message from Alice, and verifies the signature. To do this Bob needs to know Alice's public key $Pub_A$. Bob calculates the hash $e$ of the received message, scalar $u_1$ and scalar $u_2$ (see steps 2-4 in TABLE I. ). Then Bob performs two EC point multiplications and one point addition to calculate the point $T$ as shown in step 5 in TABLE I. If the x-coordinate of this point is equal to the received r, the signature was successfully verified. For the verification Bob doesn't use his own key pair $Pub_B$ and $key_B$.

## III. OUR FLEXIBLE ECDSA DESIGN

As the EC point multiplication is a complex and relatively slow operation, it is often implemented in hardware to accelerate the ECDSA operations. Our ECDSA design is an accelerator for EC point operations for the two NIST ECs *B-233* and *B-283*. The functionality of our flexible ECDSA accelerator is shown schematically in Fig. 1.

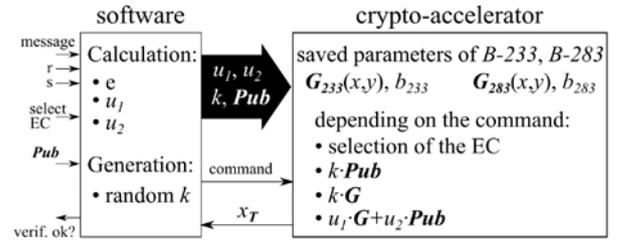

Fig. 1. Functionality of our flexible ECDSA accelerator.

Depending on the received command our accelerator selects one of the two ECs – *B-233* or *B-283* – and performs either a single $kP$ operation, for example a $k \cdot G$ multiplication for a signature generation, or multiple EC point multiplication $u_1 \cdot G + u_2 \cdot Pub$ for a signature verification. Here $G$ is the base point of the selected EC and $Pub$ is the public key used for the signature verification. The main part of our ECDSA design is a flexible accelerator for $kP$ operations.

Due to the fact that the Montgomery $kP$ algorithm is reported in literature as resistant against simple SCA attacks it is the most often one implemented in hardware. For binary ECs the most popular algorithm is the Montgomery $kP$ algorithm using Lopez-Dahab projective coordinates [18], see Algorithm 1.

---

**Algorithm 1**: Montgomery $kP$ using projective Lopez-Dahab coordinates

Input: $k = (k_{l-1} ... k_1 k_0)_2$ with $k_{l-1} = 1$, $P=(x,y)$ is a point of EC over $GF(2^l)$
Output: $kP = (x_1, y_1)$
1: $X_1 \leftarrow x$, $Z_1 \leftarrow 1$, $X_2 \leftarrow x^4+b$, $Z_2 \leftarrow x^2$
2: **for** $i=l-2$ **downto** $0$ **do**
3:   **if** $k_i=1$
4:     $T \leftarrow Z_1$, $Z_1 \leftarrow (X_1Z_2+X_2T)^2$, $X_1 \leftarrow xZ_1+X_1X_2TZ_2$
5:     $T \leftarrow X_2$, $X_2 \leftarrow T^4+bZ_2^4$, $Z_2 \leftarrow T^2Z_2^2$
6:   **else**
7:     $T \leftarrow Z_2$, $Z_2 \leftarrow (X_2Z_1+X_1T)^2$, $X_2 \leftarrow xZ_2+X_1X_2TZ_1$
8:     $T \leftarrow X_1$, $X_1 \leftarrow T^4+bZ_1^4$, $Z_1 \leftarrow T^2Z_1^2$
9:   **end if**
10: **end for**
11: $x_1 \leftarrow X_1/Z_1$
12: $y_1 \leftarrow y+(x+x_1)[(X_1+xZ_1)(X_2+xZ_2)+(x^2+y)(Z_1Z_2)] / (xZ_1Z_2)$
13: **return** $(x_1, y_1)$

This algorithm is a very efficient bitwise processing of the scalar *k*. Only 6 field multiplications, 5 field squarings and 3 field additions are necessary to process a bit of the scalar *k*, whereby the sequence of these operations is independent of the processed bit value. Authors of [19] modified this algorithm slightly to increase its resistance against SCA attacks. When implementing ECC in hardware area-optimized solutions are preferred due to manufacturing costs. Thus, only one field multiplier is usually implemented because the field multiplication is the most complex operation and the multiplier has a large area. A field squaring can be calculated as a field multiplication. In this case it doesn't require additional area but the processing of each bit of the scalar *k* consists of 6+5=11 field multiplications that have to be performed serially. Using EC over extended binary fields $GF(2^l)$ offers the possibility to accelerate the calculation of the *kP* operation significantly if the field squaring is implemented as follows:

$$C(t) = (A(t))^2 \bmod f(t), \text{ with :}$$
$$f(t) = t^{233} + t^{74} + 1 \quad \text{for } B\text{-}233;$$
$$f(t) = t^{283} + t^{12} + t^7 + t^5 + 1 \quad \text{for } B\text{-}283; \quad (1)$$
$$(A(t))^2 = (a_{l-1}a_{l-2}a_{l-3}\ldots a_1 a_0)^2 = a_{l-1}0a_{l-2}0a_{l-3}0\ldots 0a_1 0a_0$$

If the squaring is implemented corresponding to formula (1) its area is small, the calculation requires only 1 clock cycle and can be done in parallel to multiplications. Due to these facts we implemented the squaring as an additional block in our design.

Addition of $GF(2^l)$ elements is a bitwise XOR operation. Fig. 2 shows the structure of our flexible *kP* accelerator.

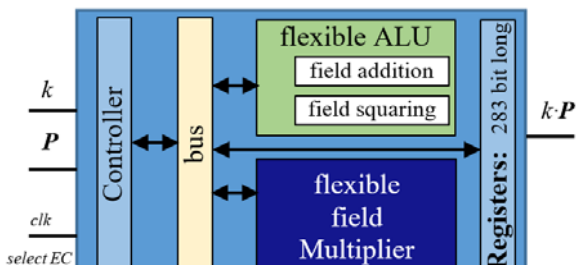

Fig. 2. Structure of our flexible *kP* accelerator.

Our design uses only one field multiplier that calculates the field product of elements of $GF(2^{233})$ for EC *B-233* or elements of $GF(2^{283})$ for EC *B-283*. Fig. 3 shows the structure of our flexible field multiplier.

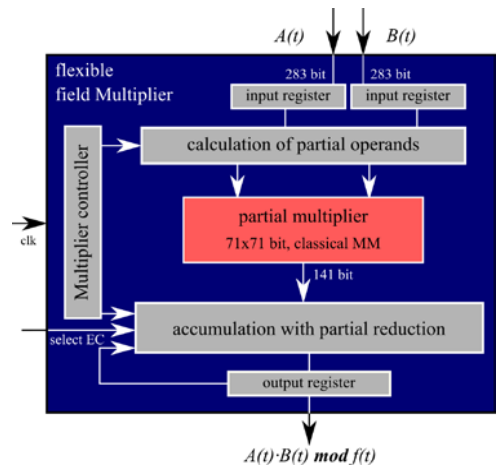

Fig. 3. Structure of our flexible field multiplier.

The polynomial multiplication (i.e. the first step of the multiplication of elements of $GF(2^l)$) can be realized by applying the classical multiplication method. Its gate complexity can be given as a number of Boolean AND and XOR operations, i.e. as the number of used AND and XOR gates. To implement the multiplication of *n*-bit long polynomials using the classical multiplication method $n^2$ AND and $(n-1)^2$ XOR gates are necessary. This results in an expensive implementation with respect to area and energy. We implemented our field multiplier using the 4-segment Karatsuba multiplication method according to a fixed calculation plan corresponding to [20]. Two up to 283 bit long operands *A(t)* and *B(t)* are segmented into four parts: $A_3, A_2, A_1, A_0$ and $B_3, B_2, B_1, B_0$ respectively. The field multiplier takes 9 clock cycles to calculate a field product. In each clock cycle only one of 9 partial products of the up to 71 bit long operands is calculated according to the 4-segment Karatsuba formula. We implemented our partial multiplier using the classical MM due to the fact that it can increase the inherent resistance of the *kP* design against selected SCA attacks[1]. All partial products are accumulated in a register of the multiplier. The field product will be accumulated iteratively, in each clock cycle, using the calculated partial products. The reduction is also performed in each clock cycle.

We ported our design implemented in VHDL to a Xilinx Spartan-6 FPGA [21] using the Xilinx ISE Design Suite version 14.7 (nt64) Parameters of our ECDSA hardware accelerator and its flexible field multiplier are given in TABLE II.

In the next section we explain how we performed the Horizontal Collision Correlation Analysis attacks against the FPGA implementation of our ECDSA design.

---

[1] If the partial multiplier is implemented corresponding to the classical MM, its area is the biggest one compared to other MMs, as well as its power consumption and the fluctuation of the power. Thus, the partial multiplier can be regarded as a kind of noise source.

TABLE II. PARAMETERS OF OUR FLEXIBLE ECDSA DESIGN

|  | Flexible ECDSA | Flexible filed multiplier only |
|---|---|---|
| **Number of Slice Registers** (out of 54576) | 6370 | 2634 |
| Number used as Flip Flops | 6369 | 2631 |
| Number used as Latches | 1 | 3 |
| **Number of Slice LUTs** (out of 27288) | 10212 | 5445 |
| Number used as logic | 10160 | 5411 |
| Number used as Memory (out of 6408) | 10 | |
| Number used exclusively as route-thrus | 42 | 24 |
| Number of occupied Slices (out of 6822) | 3559 | 1612 |
| **Number of LUT Flip Flop pairs used** | 11714 | 5674 |

## IV. ATTACK DESCRIPTION

ECDSA implementations are vulnerable to SCA attacks. If an attacker can reveal the random scalar $k$ used for a signature generation (see step 4 in TABLE I.) he can calculate the private key of Alice as follows: $key_A = (s \cdot k - e)/r \mod \varepsilon$. Finally, the attacker can examine the correctness of the revealed private key of Alice using the EC point multiplication performed for the generation of Alice's key pair ($key_A$, $Pub_A$): $key_A \cdot G = Pub_A$. This demonstrates that the implementations of EC point multiplication $k \cdot G$ have to be resistant against SCA attacks.

Usually investigations related to differential SCA attacks against ECC implementations describe classical, i.e. vertical attacks that need a lot of measurements to be performed. Horizontal attacks need by far less measurements, time and computations for revealing the key than vertical attacks. Similar to the simple power analysis horizontal DPA attacks need only one measured trace for analysis, which is performed using statistical methods. Horizontal DEMA attacks are even more dangerous as they don't require difficult preparations such as implantation of a probe resistor into the attacked board.

### A. Measurement setup

For running experiments we designed our own board for the measurement of power and especially electromagnetic traces. It is equipped with a Spartan-6 FPGA from Xilinx. We measured the current through the FPGA during a $kP$ execution. The measured current traces are the power traces. The measurement were done using the Riscure current probe [22]. For evaluation of the flexible field multiplier we measured not only the power traces but also the electromagnetic traces during the execution of field multiplications. The electromagnetic traces were measured using an MFA-R-75 EM probe from Langer [28] with an integrated amplifier. Traces were captured using a LeCroy Waverunner 610Zi oscilloscope with a 2.5 GS/s sampling rate, i.e. with 625 measurement points per clock cycle due to the clock frequency of 4 MHz. Fig. 4 shows our measurement setup and especially a zoom in for the EM measurements.

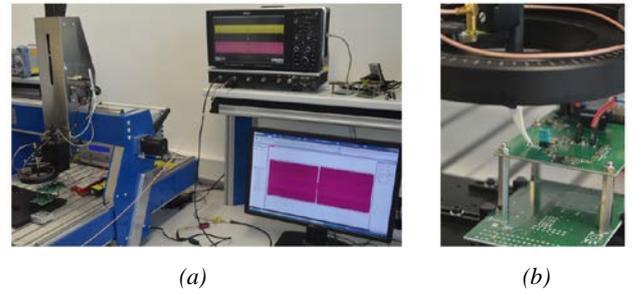

Fig. 4. Measurement setup with Langer FLS 106 IC-scanner *(a)*; PCB with the attacked FPGA and EM probe are zoomed in *(b)*.

### B. Horizontal Collision Correlation Attack

In [13] a Horizontal Collision Correlation Attack on ECs was published. The main assumption is that an attacker can distinguish two multiplications with at least one common multiplicand from two multiplications with different multiplicands. This knowledge can be used for revealing the key because a point doubling can be distinguished from a point addition even in double-and-add $kP$ algorithms obeying atomicity principle [23]. The field multiplier attacked in [13] was implemented using the classical multiplication formula. Experimental results in [24] show that Pearson's coefficients calculated for traces of two multiplications with a common operand differ significantly from coefficients calculated for two multiplications with different operands. Using this type of attack against the Montgomery $kP$ algorithm for binary ECs the scalar $k$ will not be revealed but this attack can help to separate measured traces into slots. A part of a trace that corresponds to the processing of a single key bit of the scalar $k$ is denoted as a slot. The separation of traces into slots is a preparation step of horizontal DPA and DEMA attacks [ref].

For the Montgomery $kP$ algorithm the separation of traces into slots can be done due to the fact that a multiplication with EC parameter $b$ (or a multiplication with $x$ coordinate of input point $P$) is executed in each slot in the main loop of Algorithm 1. In our implementation each slot consists of 6 field multiplications. The multiplication with the parameter $b$ is the $3^{rd}$ of the 6 multiplications. The multiplication with the $x$ coordinate of input point $P$ is the $5^{th}$ of the 6 multiplications. All other multiplications, i.e. the $1^{st}$, $2^{nd}$, $4^{th}$ and $6^{th}$ of the 6 main loop multiplications have always different operands. We performed a horizontal collision correlation attack as follows:

1 – We selected the part of the measured trace to be analysed. This part corresponds to the processing of the data in the main loop of the implemented $kP$ algorithm.

2 – We represented each clock cycle in the analysed part of the traces using only one value that we calculated using all measured values (samples) within the clock cycle, i.e. we compressed the measured trace as follows:

$$v^{compressed} = \frac{1}{N} \cdot \sum_{i=1}^{N} \left(v_i^{measured}\right)^2 \qquad (2)$$

Here $v^{compressed}$ is the averaged squared amplitude value of all samples in a clock cycle; $N$ is the number of measured values within the clock cycle, in our measurement setup $N$=625. The squaring in (2) leads to the fact that the impact of the noise in our experiments is reduced. After compressing each slot consists of 54 values and each field multiplication consists of only 9 values.

3 – We calculated an average power profile of the multiplication with the parameter $b$ using profiles of all multiplications with this operand within $kP$, i.e. we calculated the average profile of the 3$^{rd}$ of 6 multiplications in the main loop, i.e. of the clock cycles 19-27. We denote this profile as $\overline{b \cdot M2}$. This profile consists of 9 values.

4 – We calculated Pearson's coefficients for the averaged power profile denoted as $\overline{b \cdot M2}$ and each multiplication during the processing of the scalar $k$ in the main loop of implemented algorithm.

5 – We represented the calculated coefficients graphically, see Fig. 5-(a). The analysed trace contains about 1400 field multiplications performed in the main loop of the implemented algorithm for *B-233* (see left graphic in Fig. 5-(a)) and about 1700 multiplications for *B-283* (see right graphic in Fig. 5-(a)). The Pearson's coefficients calculated for the averaged power profile $\overline{b \cdot M2}$ and each multiplication with the operand $b$ are shown as red dots, all others are blue.

6 – We calculated an average power profile of the multiplication with the *x*-coordinate, i.e. we calculated the average profile of the 5$^{th}$ of the 6 multiplications in the main loop (clock cycles 37-45) and we denoted this profile as $\overline{P_x \cdot M2}$.

7 – We calculated Pearson's coefficients for the averaged power profile $\overline{P_x \cdot M2}$ and each multiplication within the processing of the scalar $k$ in the main loop of the implemented algorithm.

8 – The calculated coefficients are represented in Fig. 5-(b): the left graphic shows the results for *B-233* and the right graphic for *B-283*. The Pearson's coefficients calculated for the averaged power profile $\overline{P_x \cdot M2}$ and each multiplication with the operand $P_x$ are marked as red dots, all others are blue.

9 – Additionally, we calculated an average power profile of the 1$^{st}$ of 6 multiplications in the main loop, i.e. clock cycles 1-9. We denoted this profile as $\overline{M1 \cdot M2}$.

10 – We calculated Pearson's coefficients for the averaged power profile $\overline{M1 \cdot M2}$ and each multiplication during the processing of the scalar $k$ in the main loop of implemented algorithm.

11 – The calculated coefficients are given in Fig. 5-(c): the left side shows the results for *B-233* and the right side for *B-283*. The Pearson's coefficients calculated for the averaged power profile $\overline{M1 \cdot M2}$ and the 1$^{st}$ multiplication in each main loop iteration are marked in red, all others are blue.

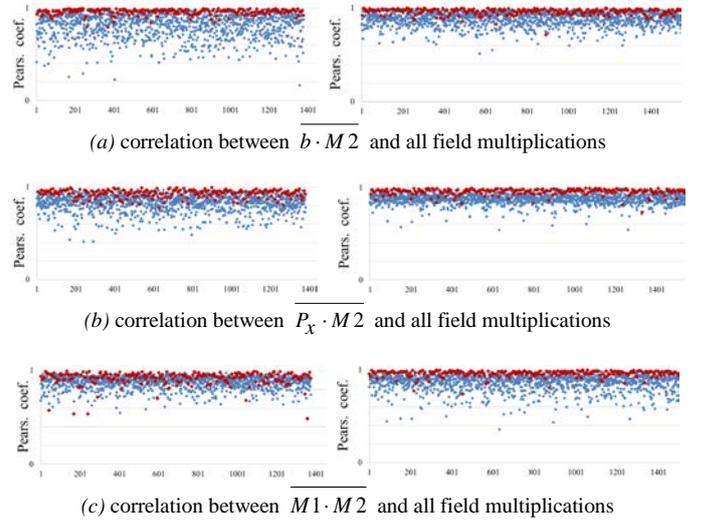

*(a)* correlation between $\overline{b \cdot M2}$ and all field multiplications

*(b)* correlation between $\overline{P_x \cdot M2}$ and all field multiplications

*(c)* correlation between $\overline{M1 \cdot M2}$ and all field multiplications

Fig. 5. Result of the horizontal collision correlation analysis: if a trace of the whole *kP* design is analysed the calculated Pearson's coefficients allow to distinguish many of the investigated multiplications (marked as red dots) from other multiplications (marked in blue).

The result of the analysis is: if a trace of the whole *kP* design is analysed the calculated Pearson's coefficients allow to distinguish many but not each of multiplications with an identical operand (see red dots in Fig. 5) from multiplications with different operands (see blue dots in Fig. 5). We assumed that not the field multiplier but other operations performed in parallel to the multiplication cause this effect. To proof this idea we analysed the power and electromagnetic traces of our field multiplier only.

V. INHERENT RESISTANCE OF FLEXIBLE FIELD MULTIPLIER

To evaluate the inherent resistance of our flexible field multiplier we implemented it without any countermeasures against SCA, i.e. the sequence of the multiplication was not randomized as for example in [25], and multiplications are not masked. We ported only our flexible multiplier to the FPGA. To investigate its resistance against Horizontal Collision Correlation Attack introduced by A. Bauer et. al. in [13] we performed the following experiments:

1. We measured the power traces of the product calculation for 4 multiplications with one common and two completely different operands: $mult_1=a \cdot b$; $mult_2=c \cdot d$; $mult_3=a \cdot e$; $mult_4=f \cdot g$.

2. We calculated Pearson coefficients $K_i$ ($1 \leq i \leq 4$) using the power shape of following multiplications: $K_1$ using $mult_1$ and $mult_3$, here operand $a$ is common in two multiplications; $K_2$ using $mult_2$ and $mult_4$, $K_3$ using $mult_1$ and $mult_2$ and $K_4$ using $mult_1$ and $mult_4$. Coefficients $K_2$, $K_3$ and $K_4$ correspond to multiplications with different operands.

If the coefficient $K_1$ differs significantly from $K_3$ and/or $K_4$, the multiplications with a common operand $mult_1$ and $mult_3$ are distinguishable from multiplications with different operands such as $mult_1$ and $mult_2$ and/or $mult_1$ and $mult_4$. Please note that this distinguishability has to be observed for each experiment (see steps 1) and 2)) for successfully revealing scalar $k$ performing an HCCA against double and add algorithms obeying the atomicity principle.

We performed the experiment described above 20 times for multiplicands of a length of 233 bits (for EC *B-233*) and of 283 bits (for EC *B-283*). Fig. 6 shows coefficients $K_1 - K_4$ calculated for all experiments. The coefficients $K_1$ are represented by red dots, $K_2$ by blue dots, $K_3$ by blue triangles and $K_4$ by blue crosses. It can be seen that the set of coefficients represented by red points is indistinguishable from other coefficients.

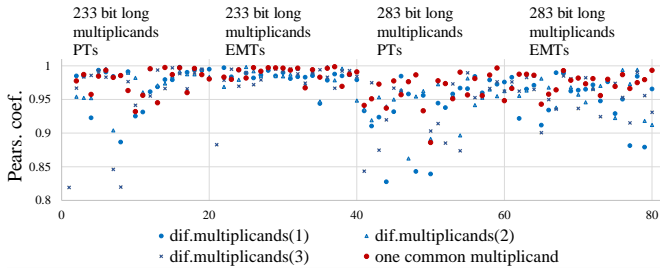

Fig. 6. Correlation coefficients coefficients $K_1 - K_4$ calculated for all experiments using power and electromagnetc traces of our flexible field multiplier: coefficients $K_1$ are marked as red dots; coefficients $K_2 - K_4$ are marked blue.

Due to the fact that the calculated Pearson coefficients do not allow to separate two groups of multiplication (see Fig. 6) i.e. multiplications with common operands cannot be distinguished from those with different operands, the inherent resistance of our investigated flexible multiplier against HCCA is high. This fact can be exploited to design a unified field multiplier, for ECs *P-224, P-256, B-233* and *B-283*. To the best of our knowledge such a filed multiplier, based on the 4-segment Karatsuba multiplication formula, was not yet reported in literature. In comparison to a multiplier based on the classical multiplication formula, the Karatsuba-based unified multiplier will be significantly faster (9 clock cycles in comparison to 16 needed by a multiplier implementing the classical MM using the same segmentation of operands), and it is inherently resistant against horizontal attacks.

## VI. CONCLUSION

In this paper we have shown that ECDSA implementations are vulnerable to Horizontal Collision Correlation Attacks. This type of attack can help to separate measured traces into slots when running attacks against the Montgomery *kP* algorithm for binary ECs. So, it is a preparation step of horizontal DPA and DEMA attacks [ref]. In order to identify the leakage source we analysed the vulnerability of the largest and most complex component of our design, i.e. its field multiplier. The result of this investigation is that the multiplier itself is resistant against HCCA. We checked this feature for two different ECs i.e. *B-233* and *B-283*. The resistance of our multiplier is due to the fact that we use the Karatsuba multiplication method with 4 segments instead of the classical multiplication method. This leads to large operands and less partial multiplications, whereby all partial multiplications are performed with different operands.

Note that our flexible inherently resistant field multiplier can also be used as a basis for designing accelerators for Elliptic Curves over prime fields such as *P-224* and *P-256*. This means it provides a valuable tamper proof-building block for hardware accelerators for any kind of elliptic curve as it is resistant against HCCA.

ACKNOWLEDGMENT

This research has been funded by the Federal Ministry of Education and Research of Germany under grant number 03ZZ0527A.

REFERENCES

[1] Federal Information Processing Standard (FIPS) 186-4, Digital Signature Standard; Request for Comments on the NIST-Recommended Elliptic Curves: 2015. DOI http://dx.doi.org/10.6028/NIST.FIPS.186-4

[2] N. Homma, Atsushi Miyamoto, Takafumi Aoki, Akashi Satoh, Adi Shamir: *Collision-based Power Analysis of Modular Exponentiation Using Chosen-message.* CHES 2008, pp. 15-29.

[3] P.-A. Fouque, F. Valette: The Doubling Attack – Why Upwards Is Better than Downwards. CHES 2003, LNCS Vol. 2779, pp. 269-280.

[4] T. S. Messerges, E. A. Dabbish, R. H. Sloan: *Power Analysis Attacks of Modular Exponentiation in Smartcards.* CHES 1999, pp. 144−157.

[5] Kocher, P., Jaffe, J., Jun, B.: Differential Power Analysis. In: Advances in Cryptology — CRYPTO' 99. pp. 388–397.

[6] Brier, E., Clavier, C., Olivier, F.: Correlation Power Analysis with a Leakage Model. In Proc. of CHES 2004. pp. 16–29.

[7] S. A. Kadir, A. Sasongko: Simple power analysis attack against elliptic curve cryptography processor on FPGA implementation, Proc. ICEEI 2011, pp. 1-4.

[8] E. De Mulder, P. Buysschaert, S. B. Ors, P. Delmotte, B. Preneel, G. Vandenbosch, I. Verbauwhede: Electromagnetic Analysis Attack on an FPGA Implementation of an Elliptic Curve Cryptosystem. EUROCON 2005, pp. 1879-1882

[9] Walter, C.D.: Sliding Windows Succumbs to Big Mac Attack. In: Proc. CHES 2001. pp. 286–299.

[10] Heyszl, J., Mangard, S., Heinz, B., Stumpf, F., Sigl, G.: Localized Electromagnetic Analysis of Cryptographic Implementations. In: Topics in Cryptology – CT-RSA 2012. pp. 231–244.

[11] Clavier, C., Feix, B., Gagnerot, G., Roussellet, M., Verneuil, V.: Horizontal correlation analysis on exponentiation.ICICS 2010, p. 46–61.

[12] Bauer, A., Jaulmes, E., Prouff, E., Wild, J.: Horizontal and Vertical Side-Channel Attacks against Secure RSA Implementations. In: Topics in Cryptology – CT-RSA 2013. pp. 1–17.

[13] Bauer, A., Jaulmes, E., Prouff, E., Wild, J.: Horizontal Collision Correlation Attack on Elliptic Curves. In: SAC 2013. pp. 553–570.

[14] Kabin, I., Dyka, Z., Kreiser, D., and Langendoerfer, P.: Horizontal Address-Bit DPA against Montgomery kP Implementation, In Proc. of the ReConFig 2017.

[15] Kabin, I., Dyka, Z., Kreiser, D., and Langendoerfer, P.: Horizontal Address-Bit DEMA against ECDSA, In Proc. of the NTMS 2018.

[16] Coron, J.-S.: Resistance Against Differential Power Analysis For Elliptic Curve Cryptosystems. CHES 1999. pp. 292–302.

[17] T. Elgamal, "A public key cryptosystem and a signature scheme based on discrete logarithms," *IEEE Transactions on Information Theory*, vol. 31, no. 4, pp. 469–472, Jul. 1985.

[18] Hankerson, D., Hernandez, J.L., Menezes, A.: Software Implementation of Elliptic Curve Cryptography over Binary Fields. In: Cryptographic Hardware and Embedded Systems — CHES 2000. pp. 1–24. Springer, Berlin, Heidelberg (2000).

[19] E. A. Bock, Z. Dyka, and P. Langendoerfer, "Increasing the Robustness of the Montgomery kP-Algorithm Against SCA by Modifying Its Initialization," in *Innovative Security Solutions for Information Technology and Communications*, 2016, pp. 167–178.

[20] Z. Dyka and P. Langendoerfer, "Area efficient hardware implementation of elliptic curve cryptography by iteratively applying Karatsuba's method," in *Design, Automation and Test in Europe*, 2005, p. 70–75 Vol. 3.
[21] Xilinx Inc.. Spartan-6 Family Overview, Product Specification. DS160 (v2.0) October 25, 2011, http://www.xilinx.com/support/documentation/data_sheets/ds160.pdf/
[22] Riscure: Inspector data sheet. Current Probe. https://www.riscure.com/benzine/documents/CurrentProbe.pdf
[23] Benoıt Chevallier-Mames, Mathieu Ciet, and Marc Joye: *Low-cost solutions for preventing simple side-channel analysis: Side-channel atomicity*, IEEE Transactions on Computersb53, No. 6/2004, p. 760-768.
[24] I. Kabin, Z. Dyka, D. Kreiser, and P. Langendoerfer, "Unified field multiplier for ECC: Inherent resistance against horizontal SCA attacks," in *2018 13th International Conference on Design Technology of Integrated Systems In Nanoscale Era (DTIS)*, 2018, pp. 1–4.
[25] F. Madlener, M. Sötttinger, and S. A. Huss, "Novel hardening techniques against differential power analysis for multiplication in GF(2n)," in *2009 International Conference on Field-Programmable Technology*, 2009, pp. 328–334.